\documentclass[a4paper]{jpconf}
\usepackage{graphicx}
\begin{document}
\title{Top 2014: Theory Summary}

\author{Tim M.P. Tait}

\address{Department of Physics and Astronomy,
University of California, Irvine, CA 92697, USA}

\ead{ttait@uci.edu}

\begin{abstract}
This write-up is the summary of the theoretical presentations at the Top 2014 Workshop held in 
Mandelieu France from September 29 to October 3, 2014.
\end{abstract}

\section{Introduction}

The 7th international workshop on Top Quark Physics was held from September 29 to October 3, 2014
in
Mandelieu, France.  I was asked to provide a summary of the 14 theoretical talks that were presented
during the workshop.  These talks were universally of very high quality, and I am afraid that the space
permitted to me does not even begin to do justice to their content.  As a result, I have had to pick
and choose some of the {\em most} striking results to present here.  There was a complementary
experimental summary which covered the remainder of the conference presentations \cite{chris}.

I organized my summary in terms of four themes:
\begin{itemize}
\item Inspiration,
\item Observation,
\item Characterization, and
\item Exploration.
\end{itemize}
This summary will follow the same plan.  I apologize in advance that most of the references will be to the write
ups each speaker provided to the conference proceedings.  I would be happy to expand these to the original
published works if any interested party cares to provide that information to me.

\section{Inspiration}

The top quark is unique among fermions in the Standard Model in that its mass is large, of order the Higgs
vacuum expectation value itself.   To provide some context, if we think of the Earth as corresponding to an
electron, then the strange quark corresponds to an object of the mass of
Jupiter and the top quark to the Sun.  This is a mystery that is crying out for some kind of explanation.
Indeed, models of flavor invariably introduce dynamics under which the top quark is either the only
``natural" quark or the only ``peculiar" one.  Either way, it is {\em the} place to look for whatever dynamics
has selected the confusing pattern of fermion masses and mixings we observe experimentally.

Even inside the Standard Model (SM), the top quark plays a unique role.  The precision with which we measure
its mass remains one of the leading uncertainties in the fit to precision electroweak data \cite{Baak:2014ora}
and the SM predictions for flavor observables \cite{Amhis:2014hma}.  In this way, it plays a primary role in
allowing these data to constrain physics beyond the SM.  
Several speakers \cite{andre,cedric}
mentioned the fact that in the Higgs potential the top mass is key
in determining whether the electroweak vacuum that we currently inhabit is the truly stable minimum,
or only a meta-stable state that will eventually decay \cite{Degrassi:2012ry}.  
In theories beyond the Standard Model
containing top partners, the large top Yukawa coupling implies that the partners
would destabilize the Higgs potential unless these partners have masses of order 
TeV \cite{cedric,yevgeny,deGouvea:2014xba}, and are thus most likely accessible at the LHC run II.\

\section{Observation}

Being now inspired to study the top quark in more detail, it makes sense to measure as many of its
properties as we can manage.  While the bulk of this hard work falls to our experimental colleagues,
theory informs and interprets these observations.  The unprecedented precision of the experimental
results demand equally precise theoretical calculations, pushing our understanding of perturbative
QCD to the limit.  Frank Krauss provided a theoretical keynote reminding us of the incredible
progress we have made in constructing Monte Carlo simulations that match the parton shower
to fixed order calculations \cite{frank}.  These advances, combined with impressive progress in
fixed order computations \cite{stefano}, are key ingredients that enable the success of the program
to make precision measurements of top quark observables.

\subsection{Top Pair Production}

Perhaps the most fundamental observation of the top quark is to identify how it is produced.
The dominant process, whereby a pair of top quarks is produced through the strong nuclear force
is currently known at next-to-next-to-leading-order (NNLO) \cite{michael}, 
allowing theoretical uncertainties on the
predicted rate to be reduced to about $5\%$.  Soft gluon re-summation further reduces the
uncertainty in the rate to about $3\%$, shared roughly equally between variation due to
the scale choice, uncertainties in the parton distribution functions (PDFs), and uncertainties in
$\alpha_S$.

These advances are not limited to the inclusive $t \bar{t}$ production rate, but also offer improved
descriptions of the kinematic distributions.  One area where this was particularly urgent was the
forward-backward asymmetry in $t \bar{t}$ production, 
\begin{equation}
A_{FB}^t \equiv \frac{N (\Delta y > 0) - N (\Delta y < 0)}{N (\Delta y > 0) + N (\Delta y < 0)}~,
\end{equation}
which famously showed a modestly
significant deviation from the SM expectation at the Tevatron,
particularly for events with larger $t \bar{t}$ invariant masses \cite{Aaltonen:2008hc,Abazov:2007ab}.
The NNLO calculation (plus leading order electroweak contributions) drives the theoretical signal
toward the data, resulting in a difference that is no longer statistically significant.  While this probably
implies that the Tevatron measurements are not revealing new physics, it is very good news for our
understanding of top production.

Despite the impressive theoretical successes in describing $t \bar{t}$ production, mysteries remain.
While the cross sections at $\sqrt{s} = 7$ and 8 TeV are each individually well-described by
the computations, the ratio of the two seems to be off by more than is expected based on the uncertainties
in the inputs \cite{alex}.  This feature is true for a variety of different PDF sets.  Another mysterious
quantity is the transverse momentum $p_T$ of the top quark itself, which shows a significant deviation
for $p_T$ greater than about $150$~GeV \cite{markus}.

\subsection{Single Top Production}

Our understanding of single top production (in the $t$-channel) has also advanced to the NNLO
level \cite{fabrizio}.  At NLO, this process has a large accidental cancellation between two subprocesses,
which could lead one to the mistaken conclusion that the uncertainties are small.  At NNLO, the process
shows variation with the choice of scale at around the percent level, 
more in line with naive expectations, and much reduced uncertainty
in differential quantities.

As one pushes single top production to higher orders of QCD, the quantum mechanical nature of
measurements asserts itself.  The various subprocesses mix with one another and also with $t \bar{t}$
production \cite{fabrizio,jan}.  This feature was observed early on in the $t W$ mode of single top 
production \cite{Tait:1999cf,Belyaev:1998dn} and argues that the language which separates these processes
can be very misleading.  A more appropriate description is to discuss a final state such as $W b W b$,
recognizing that there are resonant regions of the kinematics where diagrams containing top quarks
dominate, but also ``continuum" regions of phase space where top contributions are rather unimportant.
In intermediate regimes, the finite top width, $\Gamma_t / m_t \sim 1\%$ can play a role in precise
measurement, and tools treating the top as a final state particle (in essence, a zero-width approximation)
miss these effects.
Proper treatment can lead to significant shifts when extracting the top mass from its decay
kinematics, and very different estimates for the systematic uncertainty in the mass so extracted \cite{jan}.

\subsection{Top Polarization}

In the SM, the top decay via the electroweak interaction provides an opportunity, because the top decay
analyzes the top polarization information.  This is an important property characterizing how top quarks
are produced, and a place where one could imagine subtle manifestations of new physics
could first appear.  

One can define three observables in $t \bar{t}$ production which relate to
the expectation value of the top spin along three axes \cite{ja}.  At the Tevatron, extracting this
information is relatively easy to do, since the asymmetric $p$ and $\bar{p}$ beams provide a
natural axis one may exploit to decompose the spin information.  At the LHC, this particular handle
does not exist, but opportunities nevertheless remain.  Carefully chosen analysis can enhance the
purity of production through the $q \bar{q}$ initial state and correlate spin information with the direction
of the initial quark versus the initial anti-quark.  This could also be accomplished by considering more
favorable final states, such as $t \bar{t} \gamma$, which enhances the quark-initiated processes
compared to the less interesting gluon-initiated ones.

\section{Characterization}

Now that observation has beautifully come under control, we need to characterize the properties that we
are trying to measure.  Learning how to characterize the properties of the top quark teaches us what we
are measuring and what it means.

\subsection{The Definition of the Top Mass}

Perhaps surprisingly, the top mass itself is something whose definition is subtle.  Like any parameter in
the Lagrangian, $m_t$ is related to anything we can observe (such as the location of the
peak in the invariant mass of $Wb$) via a calculation in perturbation theory.  When working with a strictly perturbative computation, the
relation between the measurement and some suitably {\em defined} top mass can be worked out.  However,
when working with a Monte Carlo that encapsulates physics at the hadronic scale by a model, the connection
between the observable and the parameter is much less clear.  Naively, one would expect an uncertainty
caused by our ignorance of how $m_t$ is defined in the Monte Carlo of order GeV.  This is already big enough
to represent the dominant uncertainty in top mass measurements.

One proposal is to compare the output of the Monte Carlo as a function of its input mass parameter to
a calculation in soft collinear effective theory
where the the mass dependence is clear \cite{andre}.  Work has been done studying the
bottom mass dependence in $e^+ e^-$ collisions (as a warm up to the more complicated case of top
production at a hadron collider), and seems to provide a promising avenue toward relating the Monte Carlo
definition to the well-defined $\overline{\rm MS}$ mass.

\subsection{Measuring the Top Yukawa Interaction}

In the SM, the top mass relates directly to the top Yukawa coupling $y_t$ which represents 
the interaction strength of the top with the Higgs boson.  If there is physics beyond the SM, the top may
acquire mass from more than one source, and this relationship may be disrupted.  As a result, it is
of vital importance to directly measure $y_t$ independently from $m_t$.

Because the dominant production mechanism for Higgs production in the SM is through a loop of top quarks,
we already have an indirect handle on $y_t$.  However, the fact that it is indirect (with the top appearing
only virtually inside a loop) leaves open the nagging possibility that the Higgs production itself receives
contributions both from the top quark and also from new physics.  If $y_t$ is modified, but the contributions
from new physics conspire to cancel the effect of the deviation in Higgs production, we could end up fooling
ourselves that $y_t$ is standard when instead it deviates from its SM expectation.  We need a {\em direct}
measurement of $y_t$, one allowing us to establish the presence of the top quark in the process, to be sure.

Two promising techniques to search for deviations from the SM prediction for $y_t$ were 
discussed \cite{cedric}.  The first
looks at the process $t \bar{t} H$ in the decay mode $H \rightarrow \gamma \gamma$.  A high luminosity
run of the LHC at $\sqrt{s} = 14$~TeV has the eventual capability to measure $y_t$ at the $10\%$ level in
this channel.  A second proposal looks at inclusive Higgs production with a tight cut on the $p_T$ of the
Higgs.  This effectively increases the scale of the momentum running through the loop above the top mass.
If the top is the only particle running in the loop, one should see the effective interaction scale with the
momentum as the loop softens away from point-like behavior.  If there is a contribution from a heavier particle
in the loop, its contribution will continue to look like a contact interaction even for $p_t$ above the top
mass.  With high luminosity, the LHC would eventually reach sensitivity to deviations of $y_t$ of the order
of $20\%$ using this technique.

\subsection{Effective Lagrangians for Top Physics}

If there is new physics which modifies the properties of the top quark that is too heavy to be directly
observed at the LHC, it could still leave behind an influence over precise measurements.  This situation is
much like $\beta$-decay, where the interactions mediated by off-shell $W$ bosons can be observed
even at energies far below the threshold to produce the particle itself on-shell.  

The influence of such off-shell
particles manifests as a set of non-renormalizable interactions 
(often referred to as ``operators") in the effective Lagrangian.  They can
be classified as to the energy dimension of the fields involved in the term.  Their coefficients will
have units of inverse energy in order to keep the entire term mass dimension four so that the
action will be dimensionless.  The space of allowed interactions is severely constrained by the
need to respect Lorentz invariance and the SM gauge symmetries.
At low energies, lower dimension operators are more important than
higher dimension ones, because they are effectively suppressed by less powers of the process
energy.

A complete set of dimension six operators involving the top quark has been constructed, and mapped onto
observables related to the rates of top production and the properties of top decay  \cite{cen}.  
It is important
to have a complete set because at higher orders in perturbation theory, the operators mix, bleeding into
one another.  Multiple sets of operators typically contribute to any given observable, so global fits are needed to
determine the (correlated) regions of parameter space that are consistent with data.  Typical constraints
range from bounding the coupling to to be less than 
$1 / ({\rm several~hundred~GeV})^2$ to $1/({\rm a~few~TeV})^2$ (depending on the
specific interaction under consideration).  These constraints can be translated into
any theory in which new (heavy enough to be treated as contact interactions at the LHC) 
particles couple to the top quark.

\section{Exploration}

Going beyond characterizing observations, one would like to explore, seeking out fundamentally new
types of signals associated with the top quark, and understanding what they can tell us about Nature.

\subsection{Boosted Tops}

Many types of physics beyond the Standard Model result in tops that are highly boosted in the lab frame.
The prototypical examples include very massive resonances, such as the Kaluza-Klein excitations in
warped extra dimensions, which preferentially couple to top more strongly than to light fermions.  Boosted
top quarks are both a challenge and an opportunity for experimental searches.  

They are a challenge
in that they typically result in top decay products which are highly collimated, resulting in traditional
reconstructed objects which overlap in $y$-$\phi$ space and thus may not be reconstructed as top quarks
by the usual analysis designed to identify top quarks which are produced with relatively low momentum
in the lab frame.

But they are also an opportunity.  If one expects highly boosted top quarks as part of a signal, one can look
for patterns of energy distributed inside one fat jet, in order to infer the fact that it originated from a boosted
top.  Because the top decay results in a bottom quark as well as a $W$ boson's decay products, there are
features in the substructure of such a jet that are distinctive to top quark decay and allow one to distinguish
such objects from the more featureless jets produced by QCD.  There are many tools on the market
designed to accomplish this task with relatively high efficiency and purity of the resulting sample \cite{mihailo}.
Boosted topologies can help identify which reconstructed detector objects are associated with one another,
reducing combinatoric challenges that plague searches for low momentum tops.

\subsection{Tops in SUSY Decays}

As was alluded to before, a model with a top partner whose couplings to the Higgs are related to the top
Yukawa coupling demands that the mass of such a partner be close to the electroweak scale in order not
to destabilize the Higgs potential.  The most famous example of such a model is the Minimal Supersymmetric
Standard Model (MSSM), in which the scalar tops (stops) have couplings to the Higgs determined by
supersymmetry to be related to the top Yukawa.  Famously, this requires that the stop mass be not too
much larger than the electroweak scale.  Thus, if there are scalar tops in nature, they are likely to be light
and accessible at the LHC \cite{yevgeny}.

As colored particles, scalar tops should be produced at a relatively large rate at the LHC via the strong
force.  Once produced, they will decay into a (presumably lighter) neutralino together with an ordinary
top quark (which may be on- or off-shell) if $R$-parity is conserved, or into jets if $R$-parity is broken.
The majority of studies focus on the case where the theory conserves $R$-parity, due to its offered
connection to dark matter.  However, it is worth keeping in mind that supersymmetric dark matter
{\em could} turn out to be a red herring, which could leave the supersymmetric solution to the
hierarchy problem intact without the typical signals of missing transverse momentum explored
in most collider searches for supersymmetry \cite{Csaki:2011ge}.  Consequently, a plethora of signals
associated with the scalar top are possible, and a rich program of LHC physics puts bounds on the parameter
space.  Run I bounds reach several hundred GeV for many channels, and run 2 bounds are expected
to cover the bulk of natural supersymmetric theories.

\subsection{Exotic Signals of Top}

If there is new physics in the top sector, it opens the possibility for signals that one would never 
imagine in the Standard Model \cite{qh}.  For example, if there are new particles which couple to top
in a flavor-violating way, there could be ``mono-top" signals in which a single top quark is produced
in association with a particle that is sufficiently weakly interacting so as to not appreciably interact with
the detector material.  The SM rate for such a signal is vanishingly small, and its observation would be
a clear indication of physics beyond the Standard Model.

Charged scalar bosons occur in many theories of physics beyond the SM, and are likely to have enhanced
coupling to the top quark.  A charged scalar can contribute to $s$-channel 
single top production \cite{Tait:2000sh} or could be produced in association with the top through a process
such as $g b \rightarrow t H^-$.  Provided it is heavy enough, such a charged scalar would decay back
into $\bar{t} b$, producing a resonance in the $\bar{t} b$ system, as well as a polarized sample of top
quarks that help identify the nature of the charged resonance.

Another very exotic signature is production of four top quarks.  This signal is very highly suppressed
in the Standard Model, and thus there is room for substantial contributions from new physics.  In models
where the top quark is a composite object made from some more fundamental constituents, it is often
the case that this channel is the first place one might expect to realize this 
fact \cite{Lillie:2007hd,Pomarol:2008bh,Kumar:2009vs}.  Enhancements to the rate of four top production
of order $\sim 100$ remain consistent with existing data \cite{Khachatryan:2014sca}.

\subsection{Future Colliders and Top}

Precise measurements of the top quark are one of the goals of future colliders \cite{ben}.  
A high energy $e^+ e^-$
collider running above the $t \bar{t}$ threshold
can perform exquisite measurements of the top coupling to the $Z$ boson \cite{Baer:2013cma},
and even below threshold can probe the $W$-$t$-$b$ interaction to the percent level \cite{Batra:2006iq}.

At a future high energy hadron collider, exotic processes such as four top, or even six top production come
into reach.  If the new physics is described by a contact interaction (such as e.g. a chromo-magnetic
interaction), higher energy colliders enhance the
new physics contribution, since the effect of such operators grows with energy.

\section{Outlook}

Top physics is going strong, and is a model for how theory and experiment complement one another,
leading to advances that neither would be capable of alone.  It may be that the future refinement of the
current experimental analyses together with improved theoretical predictions will discover something
that does not hold together self-consistently, signaling that the Standard Model is broken.  Or it could be
that new physics will manifest itself less subtly, or even not at all.  No matter what the future holds,
the top quark will be an important part of it, and its status today defines the Standard Model and
provides an irreplaceable lens as to what kinds of new physics we can hope to discover tomorrow.

\section*{Acknowledgments}

I am grateful to the organizers of Top 2014 for the opportunity to reacquaint myself with the exciting
developments of the Top quark and the honor of preparing a summary of the conference.  
Needless to say, the excellent presentations provided by the speakers
made all of the difference in the final result.
The research of TMPT is supported in part by NSF
grant PHY-1316792 and by the University of California, Irvine through a Chancellor's fellowship.

\end{document}